%% file: 3gco.tex
\documentclass[11pt,twocolumn]{article}
\usepackage{amssymb,epsfig,wrapfig,graphics,aas_macros,appendix}
\usepackage[dvips]{color}
\usepackage[footnotesize]{caption}

\def\comment#1{{}}

\newlength{\cvindent}
\newlength{\cvhang}

\newlength{\refindent}
\begin{document}

\vspace{-0.5in}

\title{The Status and future of ground-based TeV gamma-ray astronomy\\
Reports of Individual Working Groups}
\date{}
\maketitle
\section{Galactic compact objects}
\label{GCO-subsec}
\input{3-gco.tex}

\end{document}

%% file: 3-gco.tex
Group membership: \\ \\ 
P. Kaaret, A. A. Abdo, J. Arons, M. Baring,
W. Cui, B. Dingus,
J. Finley, S. Funk, S. Heinz, B. Gaensler, A. Harding, E. Hays, J. Holder, D. Kieda, A. Konopelko, S. LeBohec, A. Levinson,
I. Moskalenko, R. Mukherjee, R. Ong, M. Pohl, K. Ragan, 
P. Slane, A. Smith, D. Torres

\subsection{Introduction}
Our Galaxy contains astrophysical systems capable of accelerating
particles to energies in excess of several tens of TeV, energies beyond
the reach of any accelerator built by humans.  What drives these
accelerators is a major question in astrophysics and understanding these
accelerators has broad implications.  TeV emission is a key diagnostic
of highly energetic particles.  Simply put, emission of a TeV photon
requires a charged particle at an energy of a TeV or greater. 
Observations in the TeV band are a sensitive probe of the highest energy
physical processes occurring in a variety of Galactic objects.  Galactic
TeV emitters also represent the sources for which we can obtain the most
detailed information on the acceleration and diffusion of high-energy
particles and are, thus, our best laboratories for understanding the
mechanisms of astrophysical ultra-relativistic accelerators.

\begin{figure*}[tb] \centerline{\psfig{file=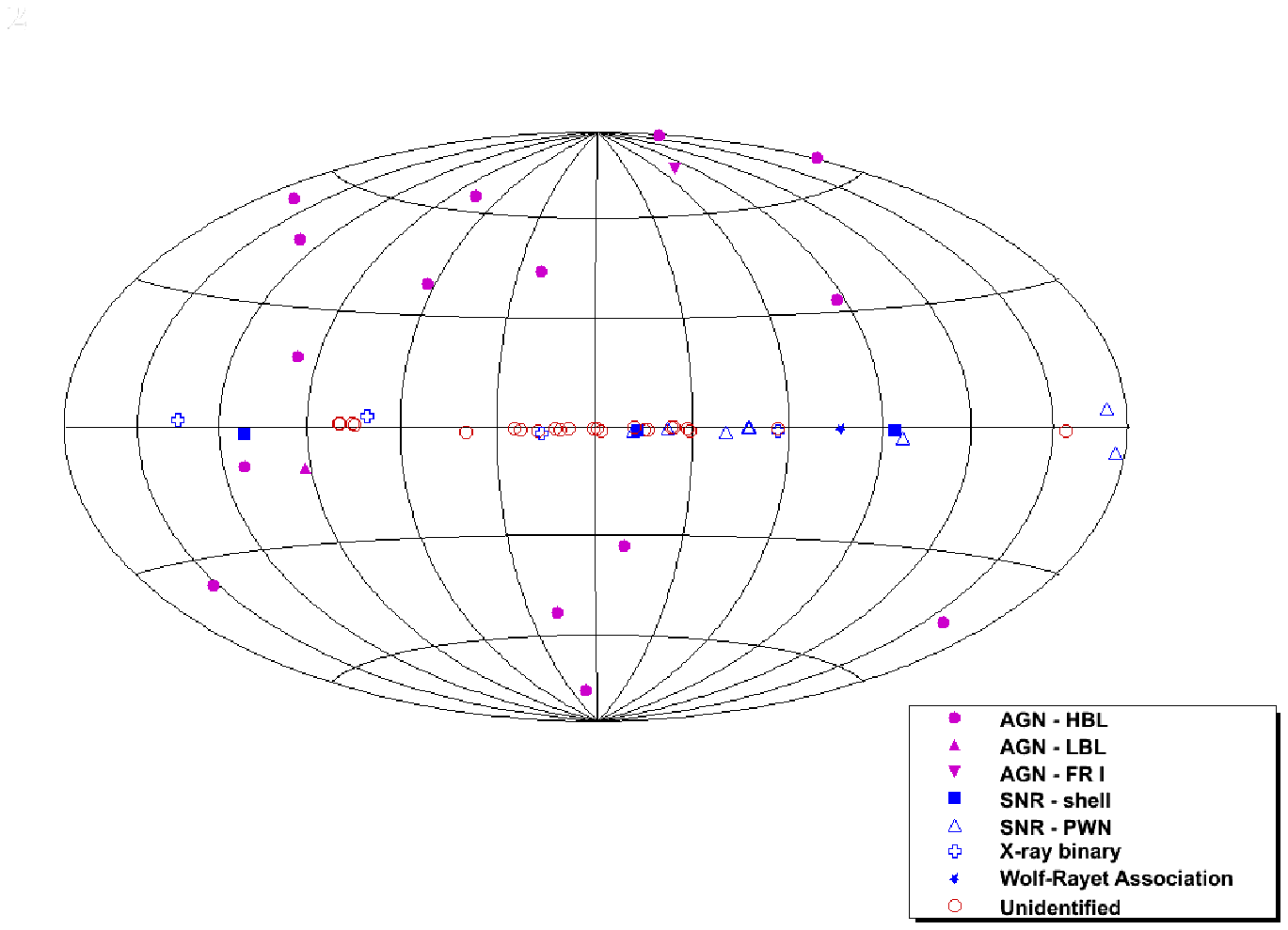,width=6.5in}}
\caption{Known TeV emitting objects plotted in Galactic coordinates. 
The center of the Milky Way is at the center of the ellipse.  The
Galactic plane is the horizontal midplane.  The symbols and colors
indicate the source type.  Figure courtesy of Dr.\ E.\ Hays.}
\label{fig:tevmap} \end{figure*}

Recent results from the new generation of TeV observations, primarily
H.E.S.S., have revealed a large population of Galactic sources; see
Fig.~\ref{fig:tevmap} which shows the known TeV sources in Galactic
coordinates.  Galactic sources now comprise a majority of the known TeV
emitters with object classes ranging from supernova remnants to X-ray
binaries to stellar associations to the unknown.  Future TeV
observations with a more sensitive telescope array will lead to the
discovery of many more TeV emitting objects and significantly advance
our understanding of the acceleration of the highest energy particles in
the Galaxy.

\subsection{Pulsar wind nebulae}

Pulsar wind nebulae (PWNe) are powered by relativistic particles
accelerated in the termination shock of the relativistic wind from a
rotation-powered pulsar.  The basic physical picture is that the
rotating magnetic field of the pulsar drives a relativistic wind.  A
termination shock forms where the internal pressure of the nebula
balances the wind ram pressure.  At the shock, particles are thermalized
and re-accelerated to Lorentz factors exceeding $10^{6}$.  The energy in
the Poynting flux is transferred, in part, to particles.  The high
energy particles then diffuse through the nebula, partially confined by
nebular magnetic fields, and cool as they age due to synchrotron losses,
producing radio to X-ray emission, and inverse-Compton losses, producing
gamma-ray emission.

Studies of PWNe address several central questions in high-energy
astrophysics, the most important of which is the mechanism of particle
acceleration in relativistic shocks.  PWNe provide a unique laboratory
for the study of relativistic shocks because the properties of the
pulsar wind are constrained by our knowledge of the pulsar and because
the details of the interaction of the relativistic wind can be imaged in
the X-ray, optical, and radio bands.  Relativistic shock acceleration is
key to many astrophysical TeV sources, and PWNe are, perhaps, the best
laboratory to understand the detailed dynamics of such shocks.  Studies
of pulsar-powered nebulae also target a number of crucial areas of
pulsar astrophysics, including the precise mechanism by which the pulsar
spin-down energy is dissipated, the ratio of magnetic to particle energy
in the pulsar wind, the electrodynamics of the magnetosphere, and the
distribution of young pulsars within the Milky Way.

Observations of TeV emission are essential to resolve these questions.
Measurement of the spectrum from the keV into the TeV range allows one
to constrain the maximum particle energy, the particle injection rate,
and the strength of the nebular magnetic field.  Observation of TeV
emission from a significant set of pulsar-powered nebulae would allow us
to study how the pulsar wind varies with pulsar properties such as
spin-down power and age.  Detection and identification of new nebulae
may also lead to the discovery of new young pulsars, particularly those
lying in dense or obscured parts of the Galaxy where radio searches are
ineffective because of dispersion.

PWNe have proven to be prolific TeV emitters.  The Crab nebula was the
first TeV source to be discovered.  H.E.S.S. has recently detected a number of other Galactic
sources, several of which are confirmed to be, and many more thought to
be, PWNe \cite{DeJager06}.  Significantly, H.E.S.S. has discovered new PWNe
that were not previously detected at other  wavelengths.  Furthermore,
the high resolution capabilities of H.E.S.S. have allowed  imaging of the
first TeV jet in the PWN of PSR1509-58 \cite{Aharonian05}, which is 
also the first astrophysical jet resolved at gamma-ray energies. 
Comparison of the  gamma-ray jet with the one detected by Chandra in
X-rays, which is less extended and  has a flatter spectral index, shows
that the evolution of emitting particles in the jet is  consistent with
synchrotron cooling.  In addition, TeV imaging has provided a clearer
picture of PWNe such as PSR B1823-13 and Vela X \cite{Aharonian06} that
are offset  from the position of the pulsar, an effect which may be due
to the pressure of the reverse  shock \cite{Blondin01}.

\subsubsection{Measurements needed}

\paragraph{Broadband modeling of PWNe}

The broadband spectrum of a PWN provides constraints on the integrated
energy injected by the pulsar as well as the effects of adiabatic
expansion and the evolution of the magnetic field.  The spectrum
consists of two components: 1) synchrotron emission extending from the
radio into the X-ray and, in some cases, the MeV band, and 2)
inverse-Compton emission producing GeV and TeV photons.  Emission in the
TeV band originates primarily from inverse-Compton scattering of ambient
soft photons with energetic electrons in the nebula.  The ratio of TeV
luminosity to pulsar spin-down power varies strongly between different
PWN and understanding the cause of this effect will advance our
understanding of the physics of PWNe.

All PWNe show spectra that are steeper in the X-ray band than in the
radio band, but the nature of the spectral changes between these bands
is not well understood.  Synchrotron losses result in a spectral break
at a frequency that depends on the age and magnetic field strength,
while other spectral features can be produced by a significant change in
the pulsar input at some epoch, by spectral features inherent in the
injection spectrum, and by interactions of the PWN with the reverse
shock from its associated supernova remnant.  TeV observations provide
an independent means to probe the electron energy distribution. 
Addition of TeV data breaks many of the degeneracies present in analysis
of synchrotron emission alone and allows independent estimates of the
electron energy distribution and the nebular magnetic field.  TeV
observations are essential to understand the PWN electron energy
distribution and its evolution.
\begin{figure}[tb] \centerline{\psfig{file=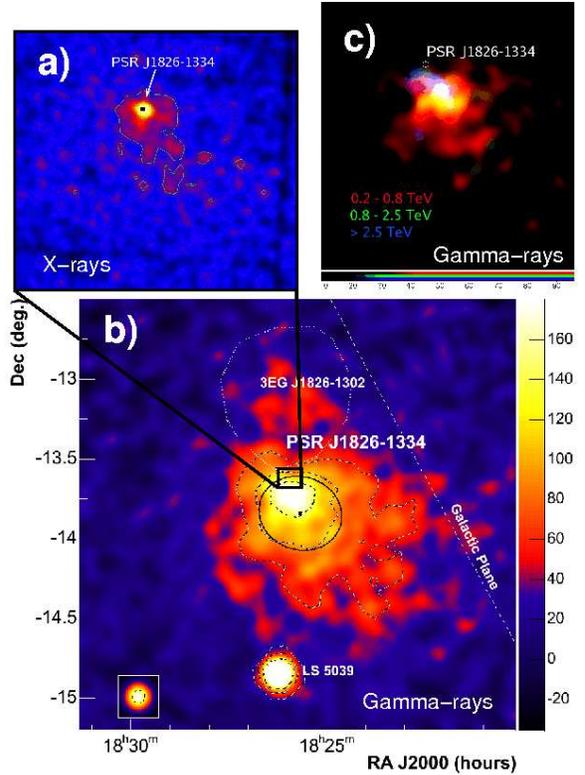,width=3.0in}}
\caption{H.E.S.S. map of TeV emission from H.E.S.S. J1825-137 (b), X-ray image
of the central part of the field showing the PWN G18.0--0.7 (a), three
color image of the TeV emission showing that the nebula is the most
compact at the highest energies (c).  From \cite{Funk06}.}
\label{fig:hess1825} \end{figure}
\paragraph{Highly Extended PWNe}
Several of the recently-discovered H.E.S.S. sources appear to be PWNe, due
to the presence of young radio pulsars nearby, but have unexpected
morphologies.  Examples include H.E.S.S. J1804-216 \cite{hess1804}, H.E.S.S.
J1825-137 \cite{hess1825}, and H.E.S.S. J1718-385 \cite{hess1718}.  There
are two issues that require considerable further study for these
sources.  First, the young pulsars suggested as the engines for these
nebulae are distinctly separated from the TeV centroids.  The most
common explanation is that the supernova remnants in which these PWNe
formed (most of which are not observed, to date) are sufficiently
evolved that the reverse shocks have disturbed the PWNe, as appears to
be the case in Vela X, which is also observed as an extended TeV source
offset from its pulsar \cite{hessVelax}.  This requires an asymmetric
interaction with the reverse shock, which can occur if the SNR expands
into a highly non-uniform medium, and there are suggestions that these
systems may indeed be evolving in the vicinity of molecular clouds.  In
this scenario, the reverse shock encounters one side of the PWNe first,
and the disruption leaves a relic nebula of particles that is
concentrated primarily on one side of the pulsar.  More sensitive TeV
observations are required to produce higher fidelity maps of these
nebulae, and to search for evidence of a steepening of the spectrum with
distance from the pulsar.

A second and more vexing question centers on the very large sizes of
these PWNe.  These sources are observed to be extended on scales as
large as $1^{\circ}$ \cite{hessVelax}, significantly larger than their
extent in X-rays.  One possible explanation for this is that the extent
of the synchrotron radiation observed in the X-ray band is confined to
the region inside the magnetic bubble of particles that is sweeping up
the ambient ejecta, while the IC emission is produced wherever energetic
particles encounter ambient photons. If the diffusion length of these
energetic particles is extremely large, they can escape the
synchrotron-emitting volume, but still produce TeV gamma rays.  Because
these sources are relatively faint, high-quality maps of this extended
emission do not yet exist. Higher sensitivity, along with somewhat
improved angular resolution, are crucial for probing more deeply into
the structure of these nebulae.
\paragraph{Jets/Magnetization}
X-ray observations with Chandra and XMM-Newton have revealed jet
structures in a large number of PWNe. Models for the formation of these
jets indicate that some fraction of the equatorial wind from the pulsar
can be redirected from its radial outflow and collimated by hoop
stresses from the inner magnetic field. The formation of these jets is
highly dependent upon the ratio of the Poynting flux to the particle
energy density in the  wind. H.E.S.S. observations of PSR B1509-58 reveal an
extended TeV jet aligned with the known X-ray jet.  New TeV observations
of similar jets should provide insight into the Poynting fraction and
the physics of jet formation.
\paragraph{Discovery Space}
The recently-discovered H.E.S.S. sources that appear to be previously
unknown PWNe highlight the potential for uncovering a large number of
PWNe in TeV surveys. For cases where the nebula magnetic field is low,
thus reducing the synchrotron emissivity, the IC emission could be the
primary observable signature. An increase in sensitivity will be
important to enhance the discovery space, and cameras with a large field
of view would enable large surveys to be conducted. Given that some of
the H.E.S.S. sources in this class are extended, improved angular resolution
also holds promise both for identifying the sources as PWNe and for
investigating the structure of these systems.

\subsection{Pulsed emission from neutron stars}

The electrodynamics of pulsars can be probed more directly via
observation of their pulsed emission.  The high timing accuracy
achievable with pulsars has led to Nobel prize winning discoveries, but
the mechanism which produces the pulsed emission, from the radio to
gamma rays, is not well understood.  TeV observations may provide key
insights.

A key question that has pervaded pulsar paradigms over the last two
decades is where is the locale of the high-energy non-thermal
magnetospheric emission?  Two competing models have been put forward for
gamma-ray pulsars: (1) polar cap scenarios, where the particle
acceleration occurs near the neutron star surface, and (2) the outer gap
picture, where this acceleration arises out near the light cylinder.
Data have not yet discriminated between these scenarios, and our
understanding of pulsar magnetopheres has stalled because of this. For
energetic young pulsars like the Crab and Vela, TeV telescopes/arrays
would offer the greatest impact if the outer gap model is operable.  For
millisecond pulsars, TeV telescopes should provide valuable insight
regardless of the emission locale.  Indeed, the answer to the question
may differ according to which subset of pulsars is examined.

Knowing the location of their radiative dissipation will permit the
identification of the pertinent physical processes involved and open up
the possibility for probing the acceleration mechanism.  This could then
enable refinement of pulsar electrodynamics studies, a difficult field
that is currently predominantly tackled via MHD and plasma simulations. 
Should polar cap environs prevail as the site for acceleration, then
there is a distinct possibility that pulsar observations could provide
the first tests of quantum electrodynamics in strong magnetic fields. 
An additional issue is to determine whether there are profound
differences in emission locales between normal pulsars and their
millisecond counterparts.  High-energy gamma-ray observations are
central to distinguishing between these competing models and accordingly
propelling various aspects of our knowledge of pulsar electrodynamics.

Detection of pulsed emission at TeV energies has so far been elusive.
The observation of high-energy cutoffs below 10 GeV in the pulsed
emission spectra of several normal pulsars with high magnetic fields by
EGRET \cite{Thompson04} has made the prospects of detecting emission at
energies above 100 GeV very unlikely. Indeed, such cutoffs are predicted
from magnetic pair production in polar cap models \cite{Daugherty96} and
from radiation reaction limits in outer gap models \cite{Romani96}.
However, outer gap models predict that a separate component produced by
inverse Compton scattering should be detectable at TeV energies, while
polar cap models do not expect such a contribution.  This provides a key
opportunity for distinguishing between these competing pictures.  Yet, the outer
gap scenario has suffered through a sequence of non-detections (e.g. 
see \cite{Lessard00} for Whipple limits on the Crab's pulsed signal)  in
focused observations by TeV telescopes, progressively pushing the pulsed
flux predictions down.  In a recent addition to this litany, MAGIC has
obtained constraining flux limits at 70 GeV and above to PSR B1951+32
\cite{magic_psr1951}, implying turnovers below around 35 GeV in the
curvature/synchrotron component, thereby mandating a revision of the
latest outer gap predictions of inverse Compton TeV fluxes
\cite{Hirotani07}.

This result highlights the importance of lowering the threshold of
ground-based ACT arrays.  Such saliency is even more palpable for the
study of millisecond pulsars (MSPs).  Polar cap model predictions can
give turnovers in the 30-70 GeV range for MSPs \cite{Harding05}, though
outer gap turnovers for MSPs are actually at lower energies due to
significant primary electron cooling by curvature radiation reaction. 
While possessing much lower magnetic fields than normal energetic young
pulsars, millisecond pulsars can be expected to be as luminous in some portion
of the gamma-ray band because their rapid periods imply large spin-down power. 
Hence, future sub-TeV observations of MSPs should significantly advance our
understanding of these objects.

\begin{figure*} \centerline{\psfig{file=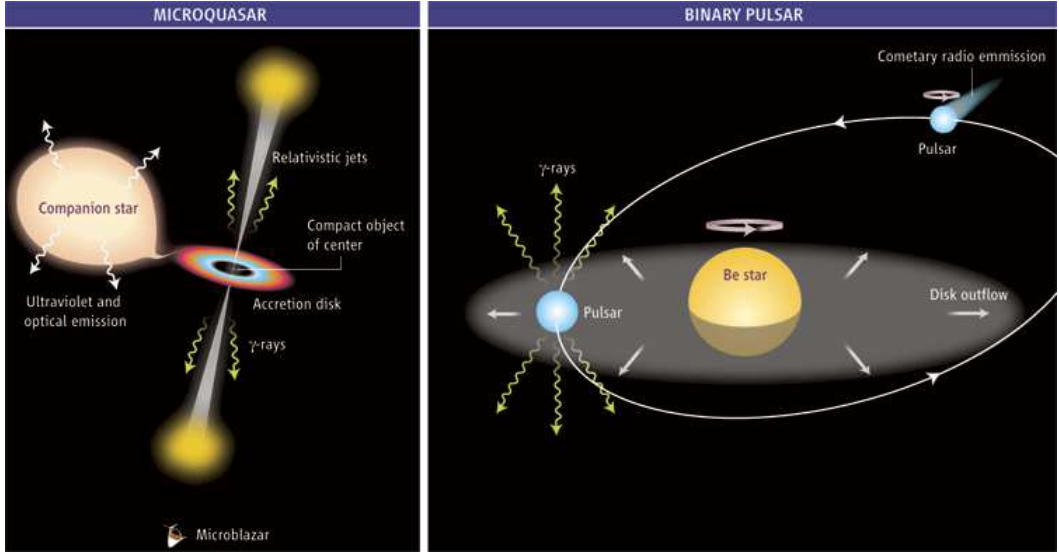,width=5.5in}}
\caption{ The two types of binaries systems producing TeV emission.  The
left image shows a microquasar powered by accretion onto a compact
object, neutron star or black hole.  The right image shows a
rotation-powered pulsar (neutron star) in a binary where the
relativistic wind from the pulsar leads to the production of TeV
photons.  From \cite{Mirabel06}.} \label{fig:binaries} \end{figure*}

\subsubsection{Measurements needed}

What is clearly needed to advance the pulsar field is a lower detection
threshold and better flux sensitivity in the sub-TeV band.  The goal of
lower thresholds is obviously to tap the potential of large fluxes from
the curvature/synchrotron component.  At the same time, greater
sensitivity can provide count rates that enable pulse-profile
determination at the EGRET level or better, which can then probe
emission region geometry. Pulse-phase spectroscopy is a necessary and
realizable goal that will enable both model discrimination and
subsequent refinement. Since the current generation ACTs cannot quite
reach thresholds below 70~GeV, and since the model predictions are very
dependent on emission and viewing geometry, it seems that detection of
very high-energy emission from millisecond and young pulsars will be
unlikely for the current instruments and will require new telescopes.
Hence, goals in the field are to both lower the threshold to the 30-50 GeV
band, and improve the flux sensitivity by a factor of ten.

\subsection[Relativistic jets from binaries]{Relativistic Jets from \\Binaries}

One of the most exciting recent discoveries in high-energy astrophysics
is the detection of TeV emission from binaries systems containing a
compact object, either a neutron star or black hole (see
Fig.~\ref{fig:binaries}).  TeV emission requires particles at TeV or
higher energies and promises to give unique insights into the
acceleration of ultrarelativistic particles in X-ray binaries.  The TeV
emission is found to be strongly time varying.  Hence, multiwavelength
(TeV, GeV, X-ray, optical, and radio) light curves will strongly
constrain models of high-energy particle acceleration and interaction
within these systems.

Key questions that will be addressed by TeV observations include:

$\bullet$~ What is the composition of ultra-relativistic jets?  Even though
ultra-relativistic jets are ubiquitous features of compact objects,
occurring in systems ranging from supermassive black holes to neutron
stars, the basic question of whether the jets are electron-positron or
have a significant hadronic component remains unanswered for almost all
objects.  The only case with a clear signature of the composition is SS
433, in which X-ray line emission reveals the presence of iron nuclei. 
However, even for SS 433, the matter may be entrained from the companion
star wind.  This question is fundamental in understanding the physics of
jet production.  Measurement of the time variation of the TeV/GeV/X-ray
spectrum from TeV emitting binaries has the potential to resolve this
question.

$\bullet$~ What is the total energy carried by jets?  TeV emission provides
a unique probe of the highest energy particles in a jet.  These
particles often dominate the total energy of the jet and their accurate
measurement is essential in understanding the energetics of jets.

$\bullet$~ What accelerates particles in jets?  Measuring the
acceleration time and the spectrum of the highest energy particles in a
jet is critical for addressing this question.

\subsubsection[Current status]{Current Status}

The first evidence that binary systems containing stellar-mass compact
objects could accelerate particles to TeV energies came from
observations of X-ray synchrotron radiation from the large-scale jets of
XTE J1550-564 \cite{Corbel02,Kaaret03}.  The detection of deceleration
in these jets suggests that the high-energy particles are accelerated by
shocks formed by the collision of the jet with the interstellar medium. 
The acceleration is likely powered by the bulk motion of the jets. More
recently, three TeV-emitting compact-object binaries have been found at
high confidence.  One, PSR B1259-63 contains a young, rotation-powered
pulsar \cite{hess1259}.  The nature of the other two systems, LS 5039
and LS I 61 303 \cite{magic_ls61} is less clear.  A lower significance
signal (3.2$\sigma$ after trials) has been reported from the black hole
X-ray binary Cyg X-1 \cite{magic_cygx1}.

PSR B1259-63 consists of a young, highly energetic pulsar in a highly
eccentric, 4.3 year orbit around a luminous Be star.  At periastron the
pulsar passes within about 1 A.U.\ of its companion star.  Radio and
hard X-ray emission, interpreted as synchrotron radiation, from the
source suggest that electrons are accelerated to relativistic energies,
mostly likely by shocks produced by interaction of the pulsar wind with
the outflow from the Be star \cite{Tavani96}.  However, the electron
energy and magnetic field strength cannot be determined independently
from the X-ray and radio data and alternative interpretations of the
X-ray emission are possible.  H.E.S.S. detected TeV emission from PSR
B1259-63 \cite{hess1259}.  TeV emission was detected over observations
within about 80 days of periastron passage and provides unambiguous
evidence for the acceleration of particles to TeV energies. 

LS I +61 303, a high mass X-ray binary system located at $\sim$2 kpc
distance which has been a source of interest for many years due to its
periodic outbursts in radio and X-ray correlated with the $\sim$26.5 day
orbital cycle and its coincidence with a COS-B and EGRET GeV gamma-ray
source \cite{Casares05,Leahy97,Taylor96}.  MAGIC found variable TeV
emission from this source \cite{magic_ls61}.  The nature of the compact
object in LS I +61 303 is not well established.   The identification of
LS I +61 303 as a microquasar occurred in 2001 \cite{Massi01} when what
appeared to be relativistic, precessing radio jets were discovered
extending roughly 200 AU from the center of the source.  However,
recent repeated VLBI imaging of the binary shows what appears to be the
cometary tail of a pulsar wind interacting with the wind from the
companion star.  This suggests that the binary is really a pairing of a
neutron star and a Be main sequence star \cite{Dhawan06}.  The (much)
shorter orbital period of LS I +61 303, as compared to PSR B1259-63,
makes the system much more accessible for observations.  Also, the
detection of LS I +61 303 at GeV energies will enable constraints on the
modeling which are not possible for PSR B1259-63.

H.E.S.S. has detected TeV emission from the high-mass X-ray binary LS 5039
\cite{hess_ls5039}.  The TeV spectral shape varies with orbital phase.
LS 5039/RX J1826.2-1450 is a high-mass X-ray binary. Radio jets from LS
5039 have been resolved using the Very Long Baseline Array
\cite{Paredes00,Bosch05}.  This suggests that the compact object is
accreting.  Optical measurement of the binary orbit also suggests a
black hole, although the measurements do not strongly exclude a neutron
star \cite{Casares05}.

\subsubsection{Measurements needed}

It should be possible to determine the correct emission mechanism for
the TeV emission in both neutron-star and black hole binaries via
simultaneous multiwavelength (radio, X-ray, GeV, TeV) observations of
the time variable emission.  Important in this regard will be measuring
how the various emission components vary with orbital phase. The key
here is adequate cadence, which requires good sensitivity even for short
observations.  Understanding the correct emission mechanism will place
the interpretation of the TeV observations on a firm footing and allow
one to use them to make strong inferences about the jet energetics and
the populations of relativistic particles in the jets.  If the TeV
emission from a given system can be shown to arise from
interactions of relativistic protons with a stellar wind, then this
would show that the jet contains hadrons.  This would provide a major
advance in our understanding of the physics of jets.

If the jets do have a significant hadronic component, then they are
potential neutrino sources.  The calculated neutrino flux levels,
assuming a hadronic origin for the observed TeV emission, are detectable
with neutrino observatories now coming on line, such as ICECUBE
\cite{Torres06}.  The detection of neutrinos from a compact object
binary would be very exciting in opening up the field of neutrino
astronomy and would be definitive proof of a hadronic jet.

Detailed light curves will also allow us to extract information about
the interaction of the pulsar wind or black hole jet with the  outflow
from the stellar companion.  This is a very exciting possibility which
will provide a direct confrontation of magnetohydrodynamical simulations
with observation and significantly advance our understanding of
time-dependent relativistic shocks.  The knowledge gained will be
important for essentially all aspects of high-energy astrophysics. If
the broad-band spectrum of PSR B1259-63 is modeled assuming that the TeV
photons are produced by inverse-Compton interactions of photons from the
companion star with the same population of accelerated electrons
producing the synchrotron emission, then the TeV data break the
degeneracy between electron energy and magnetic field and allow the
magnetic field to be estimated to be $\sim 1$~G.  This estimate is
similar to the values predicted by magnetohydrodynamical simulations of
the pulsar wind.  Future more sensitive observation would enable
measurement of the time evolution of the magnetic field.

The detection of TeV emission from a black hole binary, perhaps already
accomplished, would have important implications.  Acceleration of
particles to TeV energies is required to produce the TeV emission.  It
is unlikely that such acceleration occurs in the accretion disk or
corona; the particle acceleration likely occurs in the jet.  The same is
not true about X-ray or hard X-ray emission.  This is significant
ambiguity about whether any X-ray/hard X-ray spectral component can be
attributed to the jet, and the strong X-ray flux from the accretion
disk complicates isolation of any jet emission.  This makes TeV emission
a unique probe of the properties of jet and observation of TeV
gamma rays from the jets of accreting stellar-mass black holes should
lead to important information about the jet production mechanism.

There are two possible mechanisms for the generation of the TeV
emission.  Electrons accelerated to very high energies may
inverse-Compton scatter photons emitted from the O6.5V companion star.
However, the radiation density from the O star companion at the position
of the compact object is very high and the radiative time scale is $\sim
300$~s.  Very rapid acceleration would be required for the electrons to
reach the high energies required in the face of such rapid energy loss. 
Instead, the TeV emission may arise from the interactions of protons
accelerated in a jet with the stellar wind.  

Even with the ambiguity between an electron versus proton mechanism for
the TeV emission, the luminosity in the TeV band indicates an extremely
powerful outflow.  For very efficient, $\sim 10$\%, conversion of bulk
motion into VHE radiation, the jet power must be comparable to X-ray
luminosity.   For more typical acceleration efficiencies at the level
of a few percent, the energy in the outflow would be several times the
X-ray luminosity.  The result has major implications for our
understanding of accretion flows near black holes.  The balance between
accretion luminosity and jet power is currently a major question in the
study of microquasars, but estimation of the total jet kinetic energy
from the observed radio luminosity is uncertain \cite{Fender03}. 
Recently, a radio/optical ring was discovered around the long-known
black hole candidate Cyg X-1 \cite{Gallo05}.  The ring is powered by a
compact jet and acts as a calorimeter allowing the total jet kinetic
energy to be determined (the energy radiated by the jet is negligible). 
The jet power is between 7\% and 100\% of the X-ray luminosity of the
system.  This implies that the jet is a significant component of the
overall energy budget of the accretion flow.  It is remarkable that a
similar inference can be made directly from the observed TeV luminosity
of LS 5039.  This suggests that additional TeV sources of black hole
jet sources will be important in understanding the balance between
accretion luminosity and jet power and the fundamental role of jet
production in accretion dynamics.

A future TeV instrument with improved sensitivity would enable
observation of sources at lower luminosities than those currently
known.  An important current question in the study of Galactic black
hole is how the ratio of jet power to X-ray luminosity varies as a
function of accretion rate.  The observed relation between X-ray and
radio flux for black holes producing compact jets \cite{Corbel00} has
been interpreted as evidence that the jet dominates the accretion flow
at low accretion rates \cite{Fender03}.  Sensitive TeV observations
should enable us to directly probe this relation; the strategy would be
to observe a black hole transient in the X-ray and TeV bands as it decays
back to quiescent after an outburst.  This would provide important
information on the nature of the accretion flow at low luminosities
which would impact the question of whether the low quiescent
luminosities of black holes are valid evidence for the existence of
event horizons and also the effect of (nearly) quiescent supermassive
black holes (such as Sgr A*) on the nuclei of galaxies.

\subsection{Required instrument performance}

For the study of PWNe, the performance drivers are improved sensitivity,
angular resolution, and extension of the spectral coverage up to
100~TeV.  In order to detect large populations of fainter sources,
improved sensitivity in the band around 1~TeV is essential.  The
properties of PWNe and the resident pulsars vary significantly and a
large sample of sources is needed to fully understand these objects and it is
essential to use them as probes of pulsar astrophysics.  Improved
angular resolution, with sufficient counting statistics to make
effective use of the resolution, is needed to accurately map the TeV
emission.  Radio and X-ray maps are available with arcseconds precision
which cannot be matched in the TeV band.  However, angular resolution
sufficient to produce multiple pixel maps of the TeV emission is
adequate to map the distribution of high-energy particles as needed to
understand their diffusion within PWNe.  Extension of the spectral
coverage up to 100~TeV would enable us to measure the spectral break and
determine the highest energies to which particles are accelerated.  This
would provide fundamental information on the physics of the acceleration
process.

Since many pulsar spectra cut off below 10~GeV, extension of the energy
range down to the lowest energies possible is important for the study of
pulsed emission.  Detection of the pulsed emission from a significant
number of pulsars will likely require sensitivity below 50~GeV. 
However, a search for the inverse Compton component predicted in outer
gap models to lie at TeV energies will provide important constraints on
models.

For the study of binaries, both neutron star and black hole, sensitivity
is the main driver in order to detect additional sources and to study
known objects with high time resolution.  A factor of ten increase in
sensitivity in the `canonical' TeV band (0.2-5~TeV) should significantly
increase the number of binaries which are detected in the TeV band,
permitting studies of how TeV emission correlates with binary
properties; i.e., spin-down power, orbital separation, and companion
star type.  This will provide insights into the mechanism which produces
the TeV photons.  

Increased sensitivity is essential to study binaries at faster cadence. 
All of the binary sources are variable and the differing time
evolution at different wavebands will likely be the key to understanding
the dynamics of particle acceleration and TeV photon production in these
systems.  In addition, the ability to monitor a given source on a daily
basis for long periods is essential to allow studies of the dependence
of the TeV flux on orbital phase.  To search for jet emission from
quiescent black holes, a  flux sensitivity $10^{-14} \rm \, erg \,
s^{-1}$ in the 0.25-4~TeV band is required.

%

